\documentclass[aps,prl,oneside,onecolumn]{revtex4-2}
\usepackage{amsmath}
\usepackage{amssymb}
\usepackage{graphicx}

\begin{document}

\title{Multimode emission of fluorinated ethylene propylene clad large diameter liquid-core lasers}
\author{Anand Dewansingh$^{1}$, Abigail Deaton$^{1}$, Cortland Bergman$^{1}$, Hengzhou Liu$^{1}$, Tracy Olin$^{2}$, and Nathan J. Dawson$^{1}$}
\thanks{Corresponding author: ndawson@floridapoly.edu}

\affiliation{$^{1}$Department of Physics, Florida Polytechnic University, Lakeland, FL 33805, USA}
\affiliation{$^{2}$Department of Chemistry, Florida Polytechnic University, Lakeland, FL 33805, USA}
%\date{\today}

\begin{abstract}
A liquid-core (LiCo) dye laser was demonstrated using Rhodamine B (RhB) dissolved in glycerol as the gain medium and fluorinated ethylene propylene (FEP) tubing as the waveguide. Photoluminescence and amplified spontaneous emission (ASE) studies identified optimal RhB concentrations of 0.1 wt.\% and 0.3 wt.\% for low-threshold laser operation. Laser emission was achieved in LiCo rods with 1/16" and 1/32" inner diameter FEP tubing, with narrower tubing providing enhanced mode confinement and spectral narrowing. The addition of cavity mirrors improved emission coherence, revealing a distinct laser mode at low pump energies with mode spacing inconsistent with a simple Fabry-P\'{e}rot cavity, indicating complex mode coupling and internal reflections. Limitations include spectral broadening and scattering-induced parasitic feedback, which suggest avenues for further optimization in waveguide materials and output coupling.
\end{abstract}

\maketitle

\section{Introduction}

Dye lasers, particularly those utilizing Rhodamine B (RhB), have been widely used in research labs due to their tunable emission properties and high fluorescence quantum yields.\cite{schaf90.01,duart90.01,duart95.01} The low-cost chromophore, RhB, exhibits strong absorption in the visible spectrum, a large stokes shift, and high fluorescence quantum yield in many environments.\cite{fikry09.01} These characteristics have made it ubiquitous for applications that require a strong photoluminescence response in a polar medium. RhB has been applied in research areas such as soft matter characterization,\cite{silva05.01} real-time molecular imaging,\cite{wei24.01} microplastics,\cite{tong21.01} dentistry,\cite{dalpi06.01} and solid-state dye lasers.\cite{lu20.01,prasa23.01}

The standard dye laser cycles a solution through a cell or stream from a jet and pumped by a pulsed laser source. There have been many modern approaches that use the same laser dyes in different laser architectures such as whispering gallery mode,\cite{tomaz17.01,reyno17.01,cao24.01} distributed feedback,\cite{komik06.01,zheny08.01,andre12.01,dawso23.01} distributed Bragg reflector,\cite{singe08.01,dawso13.01,tsuts19.01} and solid-state random lasers.\cite{ander15.01,padiy22.01,he25.01} Liquid-core (LiCo) optical fibers offer the integration of organic solutions into waveguides, which can enable efficient light confinement and strong interactions with the core medium.\cite{ross81.01,chemn23.01} Light sources with waveguiding capabilities can be fabricated from LiCo optical fibers when fluorophores are introduced into the core.\cite{vasde07.01,jakub20.01} LiCo lasers have been fabricated and demonstrated using fused silica as the cladding \cite{zhou13.01,mobin17.01}. A theoretical investigation of a LiCo fiber laser has been performed for holey fibers using which also uses fused silica as the solid material.\cite{rashi18.01} Researchers have also introduced scattering particles into the solution to create LiCo optical fiber random lasers.\cite{demat07.01} Fluorinated ethylene propylene (FEP) tubing, characterized by its chemical inertness and reasonable optical clarity, serves as an effective cladding material for multimode optical fibers.\cite{dasgu99.01,dugga03.01,knitt24.01}

In this study, we investigate the lasing characteristics of a RhB/glycerol solution as the core of a LiCo laser rod with FEP cladding. refractive index contrast between the RhB/glycerol solution and the FEP cladding ensures total internal reflection at broad angles which is essential for guiding light within the core. By varying dye concentrations and employing different core diameters, we aim to elucidate the interplay between concentration-dependent photophysical phenomena and waveguide geometry. Our findings provide insights into dye-doped LiCo laser rods in a traditional laser cavity geometry for stable laser emission.

\section{Methods}

Rhodamine B (RhB) powder was added to glycerol in $20\,$ml glass vials to obtain the desired wt.\%. The RhB was purchased from Santa Cruz Biotechnology Inc., lot A2618, with a purity of $98$\% measured by high-performance liquid chromatography (HPLC). The glycerol was bottled and sold by HiMedia Laboratories LLC, lot 0000654718. Each solution was bath sonicated for two hours at $60\,^\circ$C. Immediately following sonication, while the vial remained warm, a $10\,$ml NORM-JECT\textsuperscript{\sffamily\textregistered}  syringe with a $16$ gauge dispenser extracted $1\,$ml of solution. This step was performed while the glycerol was warm because the higher temperature significantly reduced the viscosity of the solution. The partially filled syringe was then allowed to cool back to room temperature.

\begin{figure}[t]
\centering\includegraphics[scale=0.9]{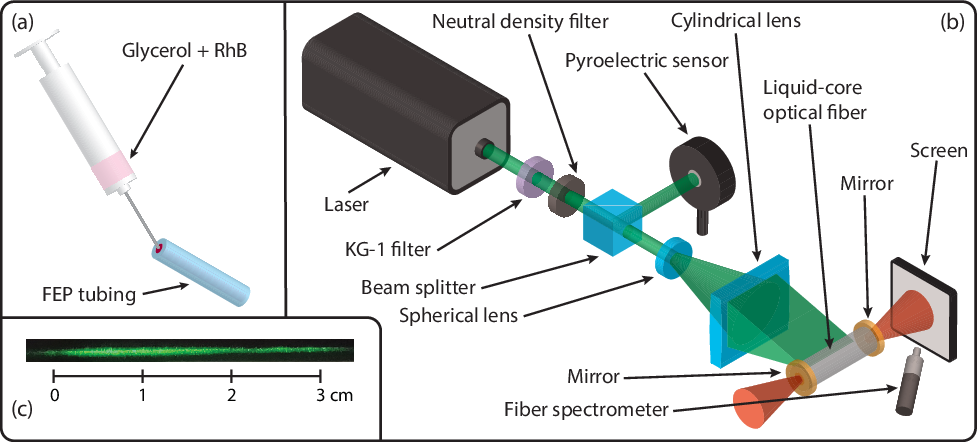}
\caption{(a) An illustration of an FEP tube being filled with dye/solvent using a syringe. (b) A diagram of the experimental setup. (c) An image of the line-focused beam used to pump the liquid-core laser rod.}
\label{fig:fig1}
\end{figure}

Fluorinated ethylene propylene (FEP) tubing with 1/16" and 1/32" inner diameters were sectioned into $3.6\,$cm segments. RhB/glycerol solutions with concentrations of $0.003$, $0.01$, $0.03$, $0.1$, and $0.3\,$wt.\% were used to fill the 1/16" inner diameter FEP tubing for amplified spontaneous emission (ASE) measurements. Based on the results from ASE experiments, segments of 1/32" inner diameter tubing were filled with high concentration, $0.1$ and $0.3\,$wt.\%, RhB/glycerol solutions. The tubes were filled by constant flow rate injection so that bubbles were not introduced into the FEP core. The procedure for introducing the solution into the FEP tubes is illustrated in Figure \ref{fig:fig1}(a). Glycerol has a relatively high viscosity; however, the low friction between the FEP and glycerol required that the tubing lay horizontal immediately after the fill.

The experimental setup shown in Figure \ref{fig:fig1}(b) characterized the emission the prepared LiCo lasers. Laser emission was recorded with cavity mirrors in place and ASE measurements were performed by removing the cavity mirrors. The pump was a frequency-doubled Nd:YAG laser emitting $532\,$nm light and operating at $10\,$Hz with $10\,$ns pulse durations. Residual $1064\,$nm light was removed with KG-1 Schott glass. Schott neutral density filters were added to adjust the pump energy incident on the samples. The beam was split with the stray beam incident on an Ophir PE10-C-BF detector to track the beam energy. The beam energy was measured at the sample location and the reference location to determine the calibration factor that determines the pulse energy incident on the sample with respect to the reference energy. The primary beam was then expanded with a spherical lens and refocused on a line incident on the LiCo rods using a cylindrical lens. Figure \ref{fig:fig1}(c) shows an image of the incident beam profile along with a scale bar. Emission from one end of the rod was projected onto a screen, where a fiber attached to an Ocean Optics HR4000 collected spectral data.

\section{Results and Discussion}

The model laser shown in Figure \ref{fig:fig1} will emit laser light when a population inversion density reaches a critical threshold. This population inversion density threshold is based on the quantum yield of the fluorophores and their concentration in the solvent. The desired population inversion density for relatively low-threshold laser emission is achieved at higher dye concentrations; however, very high dye concentrations can cause the laser threshold to increase. For example, a reduction in extrinsic photoluminescence measurements are observed at high concentrations which can be caused by both dye aggregation and reabsorption. To determine a suitable dye concentration, a series of LiCo laser rods were fabricated with 1/16" core diameter FEP tubing. The FEP tubing was intensionally chosen to have twice the diameter of the laser rod tubing because photoluminescence measurements at high dye concentrations showed reduced signs of feedback-induced spectral deformations at lower intensities relative to the same measurements using the 1/32" inner diameter tubing.

\begin{figure}[t]
\centering\includegraphics[scale=1]{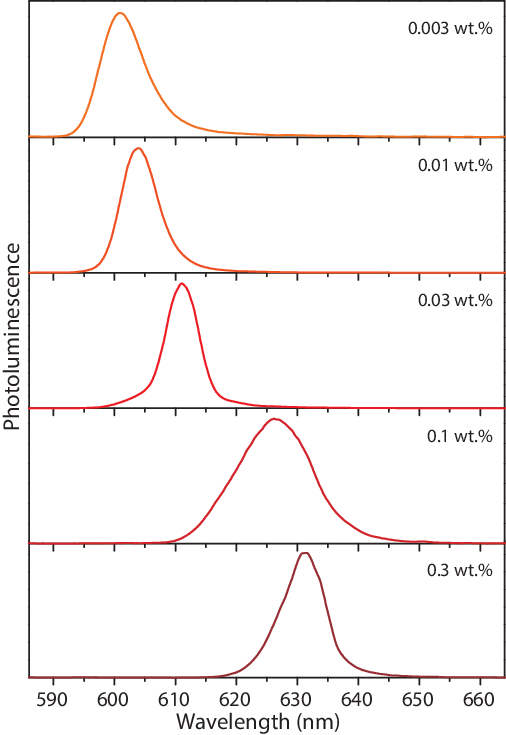}
\caption{The photoluminescence emitted from the end of a mirrorless liquid-core rod. The redshift of the peak emission and changes in the spectral profile occur as the intensity is increased.}
\label{fig:fig2}
\end{figure}

Photoluminescence measurements of RhB/glycerol-filled FEP tubing with 1/16" core diameter are shown in Figure \ref{fig:fig2} for various dye concentrations. All emission spectra were measured from the end of the fiber. There is a clear red shift in peak luminescence as the concentration increases. There were was some variation between the spectral profile of different samples at the same concentration and with slight changes in sample position relative to the pump and detector; however, all measurements showed the same large redshift in peak emission between rods filled with $0.03\,$wt.\% RhB and those filled with $0.1\,$wt.\% RhB. The dominant mechanism for the observed redshift in photoluminescence, especially in lower-concentration samples, is re-absorption. At shorter wavelengths, the absorption spectrum overlaps with the fluorescence spectrum. At higher concentrations, amplification of the spontaneous emission can become significant. Additionally the quantum yield at higher concentrations is reduced while the absorption is increased. We attempted to test a $1\,$wt.\% concentration sample, but the emission intensity was too low for further consideration as a concentration to be used in a laser in the proposed geometry.

\begin{figure}[t]
\centering\includegraphics[scale=1]{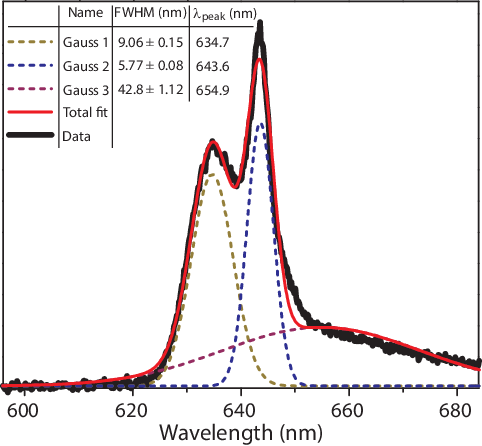}
\caption{An example of a bimodal emission spectrum occurring at RhB concentration of $0.3\,$wt.\% in glycerol. The dashed lines are Gaussian fit functions while the red line and thicker black line are the respective cumulative fit and experimental data.}
\label{fig:fig3}
\end{figure}

RhB/glycerol-filled FEP tubes with concentrations at $0.1\,$wt.\% and $0.3\,$wt.\% showed the largest uncertainties between spectral profiles, where both the full-width at half maximum (FWHM) and peak emission wavelength showed large variation. As observed in Figure \ref{fig:fig2}, the spectrum for the $0.1\,$wt.\% concentration is relatively broad as compared to the other shown spectra. The broadness of the peak is not associated with less amplification, but rather, the spectrum shows multiple modes. The sample at $0.3\,$wt.\% concentration shown in Figure \ref{fig:fig2} has a narrower spectral profile. A different sample prepared at $0.3\,$wt.\% concentration illustrates a more pronounced set of peaks as shown in Figure \ref{fig:fig3} caused by the same mechanism that broadened the $0.1\,$wt.\% concentration sample shown in Figure \ref{fig:fig2}.

The three-Gaussian fit shown in Figure \ref{fig:fig3} can capture much of the spectral features. The typical ASE curve can be fit with a single Gaussian. If a large asymmetry exists in a redshifted ASE peak emitted from long strip lines at high concentration, then the cumulative sum of two Gaussians can often be used to fit the function. Two broadest Gaussian functions shown in Figure \ref{fig:fig3} capture much of the traditional ASE spectrum. The other Gaussian fits the narrowest Gaussian function well. This narrow peak is which is attributed to enhanced ASE from non-resonant feedback caused by many modes supported in the LiCo gain medium through total internal reflections. Although not the exact same mechanism, this phenomenon has similarities to a scattering medium that produces non-resonant feedback (intensity feedback) random laser emission. Based on the results from photoluminescence measurements, the best candidates for LiCo laser rods had RhB/glycerol concentrations of $0.1\,$wt.\% and $0.3\,$wt.\%.

Two LiCo lasers were made from the $0.1\,$wt.\% solution. The same 97\% reflectance mirrors were used for these lasers. The first fabricated LiCo laser at a $0.1\,$wt.\% concentration had a 1/16" core diameter. The emission spectra for various pump powers is shown in Figure S1 of the supplementary document. A relatively broad multimode laser peak appears above threshold and centered at $\sim 630\,$nm. Figure S2 shows the emission spectra for the second fabricated LiCo laser made using the same $0.1\,$wt.\% concentration. This second LiCo laser used FEP tubing with a 1/32" inner diameter. Another broad multimode emission peak emerged above threshold; however, the individual modes are more distinguishable in the plotted spectra. The second laser was also centered at $\sim 630\,$nm

\begin{figure}[t]
\centering\includegraphics[scale=1]{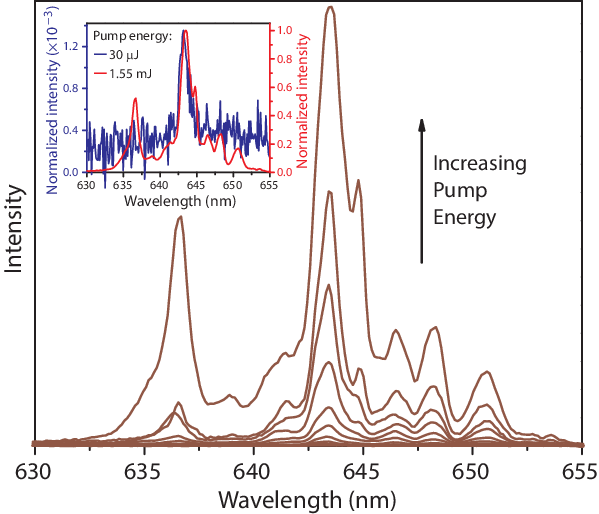}
\caption{Emission spectra showing laser emission lines for different pump energies. The inset shows the emission spectrum for the lowest and highest pump energies tested.}
\label{fig:fig4}
\end{figure}

A third laser was made from the $0.3\,$wt.\% RhB/glycerol solution. Based on the results of the $0.1\,$wt.\% lasers, the third LiCo laser was fabricated by filling 1/32" inner diameter FEP tubing with the $0.3\,$wt.\% solution. Figure \ref{fig:fig4} shows the emission spectra for the $0.3\,$wt.\% laser for various pump energies. A large single mode peak appears at relatively low pump energy, $\sim 30\,\mu$J. The left vertical axis for the inset in Figure \ref{fig:fig4} was rescaled to show the laser peak more clearly. The lack of an output coupler resulted in relatively weak emission intensities. A large spectrometer integration time of $2000\,$ms (20 pulses) was necessary to resolve the signal.

The inset of Figure \ref{fig:fig4} shows relatively low threshold laser emission, which is near the lowest detectable energy available to easily identify the signal through the spectrometer's noise. Both $12\,$cm-diameter cavity mirrors were $97$\% reflectance dielectric reflectors. The addition of a lower reflectance output coupler on one end would have yielded much stronger output emission.

\begin{figure}[t]
\centering\includegraphics[scale=0.8]{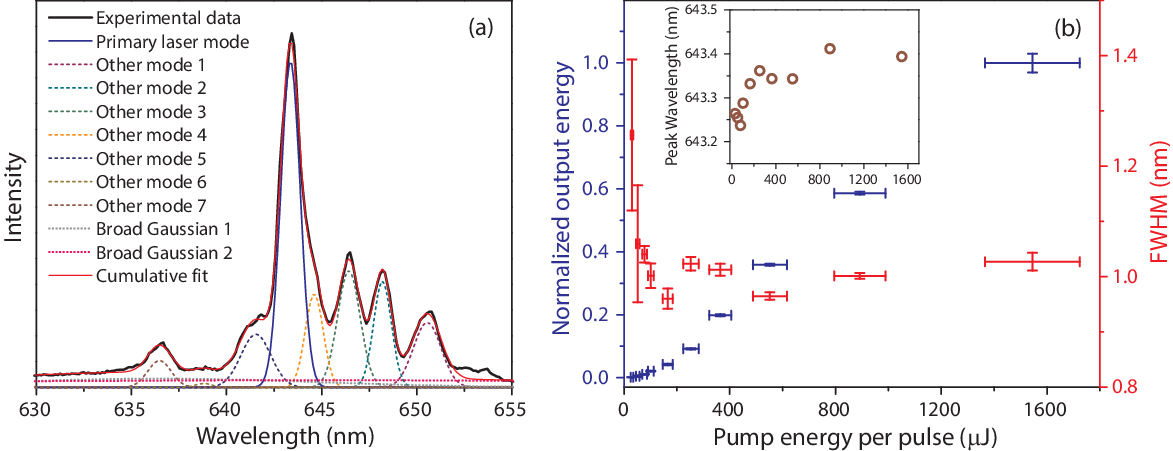}
\caption{(a) A 10-peak Gaussian fit to an emission spectrum collected from one end of $1/32$" inner diameter LiCo laser filled with $0.3\,$wt.\% RhB/glycerol. (b) The spectral width of the primary laser mode and its power dependence. The primary laser mode's peak wavelength as a function of pump power is shown in the set.}
\label{fig:fig5}
\end{figure}

Figure \ref{fig:fig5}(a) shows the emission spectrum at a pump power of $250\,\mu$J. Ten Gaussian peak functions were fit to the experimental data to reproduce the spectrum. The two broad Gaussians that fit the incoherent emission had large uncertainties due to dependencies between those two highly overlapping functions. The other peak functions easily settled into the spectrum's ``notches'' using a least squares regression algorithm. The ``Primary laser mode'' peak function shown in Figure \ref{fig:fig5}(a) along with ``Other mode 2" and ``Other mode 4" are the narrowest peak functions. The average distance of nearest-neighbor peak separation between the primary laser mode peak and the other mode peaks is $2.0\,$nm. The uncertainty in the average distances of separation can be estimated with the corrected standard deviation, which yields $0.5\,$nm.

There are likely several contributing factors to the observed comb-like emission spectrum. If we naively assume a perfect one-dimensional system, then the wavelength-dependent free spectral range (FSR) $\Delta \lambda_\mathrm{FSR}$ must be equal to the measured separation distance between modes, $\Delta \lambda = \left(2.0 \pm 0.5\right)\,$nm. Therefore, the cavity length can be determined if we can assume that $\Delta \lambda_\mathrm{FSR} = \Delta \lambda$. The frequency-dependent FSR is given by $\Delta \nu_\mathrm{FSR} = c/2nL$,\cite{saleh07.01} where $n \approx 1.47$ is the refractive index of the solution, $c$ is the speed of light in vacuum, and $L$ is the length of the ideal one-dimensional cavity. Using the relationship between frequency and free-space wavelength, $\nu = c/\lambda$, or $d\nu = - \left(c/\lambda^2\right)d\lambda$, the cavity length can be calculated, which yields $L = \lambda^2/2n\,\Delta \lambda_\mathrm{FSR} \approx \left(70 \pm 19\right)\,\mu$m. Clearly the $3.6\,$cm cavity is much longer than $70\,\mu$, which means that we cannot assume the measured $\Delta \lambda$ is associated with the $\Delta \lambda_\mathrm{FSR}$ of an ideal one-dimensional cavity architecture. Bright laser emission was also detected from the curved surface of the FEP tubing which was caused by scattering. Output from scattered light caused by a resonance associated with an inner normal reflection or a whispering gallery mode can also occur. The $1/32$" inner diameter of the FEP tubing is about $800\,\mu$m, and therefore we cannot account for a pure FSR from smooth surface reflections normal to the FEP tubing interface. The round trip associated with a whispering gallery mode is longer than trips across the diameter from normal reflections at the tubing interfaces. We also observed that the emission was much broader without the addition of cavity mirrors. Therefore, we can conclude that 1) the cavity mirrors are essential for narrow laser emission and 2) the measured $\Delta \lambda$ is not purely associated with Fabry-Perot cavity modes. The spacing between modes is relatively large in Figure \ref{fig:fig5} in relation to the spectra shown in Figures S1 and S2.

The modes shown in Figure \ref{fig:fig5}(a) are not associated with one-dimensional, Fabry-Perot, cavity transmission modes; however, these modes can be associated with the radial modes of a large diameter step-index fiber. The number of radial modes for a given wavelength in the multimode liquid core fiber can be approximated \textit{via} \cite{wentw05.01}
\begin{equation}
N \approx 2\left(\pi r/\lambda\right)\left(n_\mathrm{core}^2 - n_\mathrm{clad}^2\right) \, ,
\label{eq:radmodes}
\end{equation}
where $r$ is the inner radius of the tubing, $\lambda$ is the free-space wavelength, and $n$ denotes the index of refraction. For FEP tubing with a $\sim 800\,\mu$m diameter ($400\,\mu$m radius) and a refractive index of 1.34 at $\lambda = 640\,$nm (and $n_\mathrm{core}\approx 1.47$), we can estimate $N \sim 2.8\times 10^{6}$. With such a large number of radial modes supported by the fiber and a high density of cavity modes over the gain region of the spectrum, there appear to be a relatively low number of laser modes in Figure \ref{fig:fig5}(a). This can be explained from scattering losses at the FEP tubing interface, where high-order modes that travel closer to the cladding on average experience significantly greater cavity loss.

Let us assume that the primary laser mode is a relatively low ordered mode with light near the center of the fiber, on average. This assumption is based on the primary mode having the largest amplitude, where cavity loss from FEP scattering is minimal. We can denote the effective refractive index of the primary mode as $n_\mathrm{p}$, where we make the approximation $n_\mathrm{p} \approx n_\mathrm{core}$. From a ray optics interpretation of a multimode fiber, the internally reflected light occurs at the angle $\theta$ given by \begin{equation}
\sin \theta = \frac{\sqrt{n_\mathrm{core}^2 - n_\mathrm{eff}^2}}{n_\mathrm{core}} \, .
\label{eq:sineq}
\end{equation}
An average wavelength shift of $\sim 2.0\,$nm from the primary mode of wavelength of $\lambda \approx 643\,$nm results in a refractive index change of $\Delta n = \left(\Delta \lambda\right) n_\mathrm{p}/\lambda\approx 4.6\times 10^{-3}$. Assuming the primary mode is in the zeroth mode gives the largest possible change in angle $\Delta \theta_\mathrm{max}$ for a $2.0\,$nm shift in wavelength from an adjacent mode. From these values, we find that $\Delta \theta_{\mathrm{max}}^{2\,\mathrm{nm}} \approx 4.5^\circ$. This maximum possible angle is indeed notable and could be potentially observed without incredibly fine tuned filters, if the line widths were delta functions and there was no mode mixing from scattering within the system. The double-shifted maximum angular separation possible from the primary mode, assumed to be at the zeroth mode, for a $\sim 4\,$nm shift is $\Delta \theta_{\mathrm{max}}^{4\,\mathrm{nm}} \approx 6.4^\circ$. The angular separation of the triple-shifted operational mode follows as $\Delta \theta_{\mathrm{max}}^{6\,\mathrm{nm}} \approx 7.8^\circ$. Of course, the primary mode can be slightly angled, where additional peaks are observed at wavelengths below the primary mode. If the system was diffraction limited, then we would expect some identifiable angular dependence; however, the $3.6\,$cm cavity is rather short and the collimation exiting the laser was rather poor relative to a large solid-state lab laser. The broadness of the laser lines in relation to the separation also causes significant overlap as well, even in the high concentration case shown in Figure \ref{fig:fig4} which showed the best spectral features of all concentrations studied. The end result is a smearing of closely spaced laser lines that are not easily distinguishable using our experimental protocol which directs a $300\,\mu$m-core diameter fiber spectrometer probe at the laser spot. Modifying the experiment to scan the fiber over the emission profile could reveal some smoothed version of multimode emission lines we anticipate from a model system.

Figure \ref{fig:fig5}(b) plots the fitted values for the full-width at half-maximum (FWHM) and the relative area under the curve for the primary laser mode. Although the peak was observed at the lowest recordable intensity level, the FWHM decreased as the output intensity increased. The peak quickly narrowed to a steady FWHM value of $\sim 1\,$nm when the pump energy was increased. A change in the slope of the input/output energy curve is also observed in Figure \ref{fig:fig5}(b), where the slope efficiency appears more parabolic than the anticipated threshold behavior; however, the low-threshold nature of the emission and immediate emergence of the primary laser mode suggests a threshold that is lower than our recording capabilities. The operating wavelength of the primary laser mode is shown in the inset of Figure \ref{fig:fig5}(b). The redshift of the primary laser mode when the pump energy increases indicates a local temperature change that increases the optical path length.

There are several advantages and disadvantages with the presented LiCo laser system. The lack of sensitivity is possibly the most advantageous aspect of the design. The flat mirrors can be placed between two wetted ends of the rod and require very little tuning. For example, the ``bounce'' of photons off the FEP tubing interface from total internal reflection and the vast number of modes supported by the $1/32$" inner diameter tubing allow for newly supported laser modes to operate if the alignment of the external mirrors are slightly altered. There are, however, losses are associated with scattering from the FEP, where the lowest threshold, primary laser mode will be among the relatively lower-ordered radial modes of the fiber with a relatively high effective refractive index. This self-correcting attribute is well suited for applications where a system can experience deformations during transport and operation. This advantage is directly related to many of the disadvantages as well. For example, the large number of fiber modes and cavity modes are observed to support other modes of operations that reduce the total available energy for the primary laser mode. Additionally, the low-index FEP material is not perfectly transparent, where lossy scattering can also cause mode coupling as well as the emergence of parasitic laser emission from the cladding. Feedback from scattering in the FEP could also support random laser emission at high pump intensities which would further reduce the laser output efficiency.

The mirrors used in this study were broadband dielectric reflectors. Some past studies used fused silica to contain a much smaller diameter liquid core, where tuning was achieved via small changes in the cavity length.\cite{zhou13.01,mobin17.01} This multimodal, self-correcting behavior of the present system keeps it from being tuned in any significant way by slightly adjusting the Fabry-P\'{e}rot cavity length. In fact, one would expect to see large fringes appear as it is tuned. Again, we did not observed this behavior, and we believe that scattering losses at the FEP interface disallowed higher-ordered radial modes of the fiber from operating as a laser for the range of pump powers used in this study. Two methods are readily available for fine tuning the resonance of the primary laser mode. The first fine-tuning method is the so-called ``temperature tuning'' method, where the resonance redshifts as the temperature of the system is increased. The effects of temperature changes on the resonant wavelength from pumping alone are shown in the inset of Figure \ref{fig:fig5}(b). A second method of fine tuning that will not affect the macroscopic geometry can be achieved by tuning the reflection band of dielectric mirrors when the operating wavelength in near a band edge. For example, polymeric distributed Bragg reflector lasers are known to prefer band edge laser emission, where the sharp dispersion associated with the band edge results in the lowest group velocities amount the Fabry-P\'{e}rot cavity modes.\cite{andre14.01} Therefore, the mirrors at the end of the LiCo laser could be tailored to produce the desired emission lines over the gain profile. Furthermore, tunability of Bragg mirrors fabricated from soft materials has been demonstrated through temperature changes \cite{andre13.01} and mechanical strain.\cite{mao11.01}

The above demonstration can help finely tune the primary mode across the gain profile; however, the active material's gain profile can be tuned across the laser dye's entire inhomogeneous-broadened gain envelope based on reabsorption losses as shown in Figure \ref{fig:fig2}. Reducing/increasing the concentration will blueshift/redshift the gain profile. Likewise, shortening/lengthening the LiCo fiber will also blueshift/redshift the gain profile by decreasing/increasing losses from reabsorption. Note that the methods used for broad tuning of the gain profile will directly affect device geometry and other parameters of the active material beyond the spectral distribution.

In addition to shortcomings associated with scattering from the FEP cladding and multiple modes supported by a large-diameter fiber, the laser's gain is derived from the RhB laser dye. In general, laser dyes have high fluorescence quantum yields; however, the branching ratio of even the best laser dyes is not unity. After laser dye molecules such as RhB have been excited, there is a probability that some will not relax into the ground state, but rather follow a pathway that ultimately leads to photobleaching of the gain medium. Therefore, there is a finite lifetime proportional to the pump fluence based on the type of fluorophore used and its local environment. Kiraz \textit{et} \textit{al}. noted this photodegradation/photobleaching phenomena in the RhB/glycerol system,\cite{kiraz07.01} where photobleaching in microdroplets have been documented in other rhodamine derivative/solvent systems such as rhodamine 6g in a methanol/glycerol solvent.\cite{ifti23.01}

A small photobleaching study was performed with a $\sim 100\,$mW CW laser at $532\,$nm. The peak power of the pulsed pump is much larger than the laser diode, but the average power is quite low, where the same amount of energy of a $100\,\mu$J pulse is absorbed by the system using the CW laser in $1\,$ms. If the rate at which RhB molecules bleach during stimulated emission is similar to a system undergoing spontaneous emission and nonradiative decay only, then the system will decay the same in $1$ second of illumination by the CW laser as it would for $1000$ pulses at $100\,\mu$J/pulse. The decay fit from Figure S3 in the supplementary document indicates a decay rate of $\sim 0.12\,$Np/min, which indicates a robust system. Furthermore, the glycerol can replenish RhB molecules from dark regions. The constant replenish rate model shows an asymptote that approaches $>85$\% of the original photoluminescence signal; however, local heating due to the CW laser is greater than the pulsed source which increases the replenish rate. The diameter of the fiber and shallow penetration depth of the pump at high RhB concentration indicates that replenishment also occurs in the LiCo laser during operation, which further improves the system's resistance to photobleaching signal loss during operation. It is worth making a final note about some other strategies have been shown to improve the lifetime of chromophores such as structural changes to the molecules,\cite{lipha83.01} the addition of antioxidants,\cite{macke01.01} and encapsulation.\cite{trofy14.01,stein18.01} Encapsulation into large, index-matched, polymers that have been shown to mediate fluorophore recovery \cite{peng98.01,howel02.01,ander16.05,dhaka16.01,stubb18.01,chris19.01} might also prove useful for extending the device's lifetime.

\section{Conclusion}

A dye laser formed from a LiCo laser rod was demonstrated using RhB dissolved in glycerol as the gain medium, FEP tubing as the waveguiding structure. Various RhB concentrations were evaluated to identify optimal conditions for ASE and laser emission. ASE measurements indicated that concentrations of $0.1\,$wt.\% and $0.3\,$wt.\% provided the most favorable conditions for low-threshold laser operation, exhibiting both multimode ASE and spectrally distinct features suggestive of non-resonant feedback and mode coupling.

Laser emission was achieved in rods formed from both 1/16" and 1/32" inner diameter FEP tubing. The narrower tubing facilitated tighter mode confinement and narrower spectral features, while the addition of cavity mirrors significantly improved emission coherence. At $0.3\,$wt.\%, a prominent laser mode emerged at low pump energies, with mode spacing near $\left(2.0 \pm 0.5\right)\,$nm. This spacing is inconsistent with a simple Fabry-P\'{e}rot cavity model, implying contributions from transverse mode coupling, cavity losses, and internal reflections.

The design offers several practical advantages, including high tolerance to mechanical misalignment and minimal cavity tuning requirements. The high number of supported modes and internal reflections within the low-index FEP tubing enable lasing to persist under varying geometric conditions, making the system well-suited for portable and field-deployed applications. However, spectral broadening, mode competition, and scattering-induced parasitic feedback and laser emission reduce output efficiency. Further improvements may be achieved by incorporating optimized output coupling and selecting waveguide materials with lower scattering losses and higher transparency.

\section*{Acknowledgments}

This material is based upon work partially supported by the National Science Foundation, Directorate for Mathematical and Physical Sciences (Grant No. 2337595). This work was also partially supported by Florida Polytechnic University Technology Fee funds.

%\bibliographystyle{iopart-num}
%\bibliography{LiCoBib}

\providecommand{\newblock}{}

\clearpage

\section*{Supplementary documentation}

\setcounter{figure}{0}        % restart numbering
\renewcommand{\thefigure}{S\arabic{figure}}  % add "S" prefix

\noindent This section contains additional graphical data in support of the main article.

\begin{figure}[!h]
\centering\includegraphics[scale=0.9]{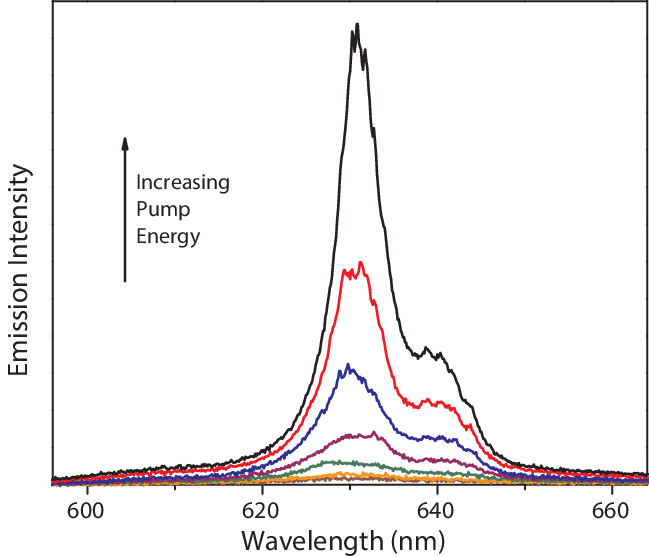}
\caption{Emission spectra from a LiCo laser made by filling FEP tubing with an inner diameter of 1/16" inch with a $0.1\,$wt.\% RhB/glycerol solution. The reflectors on each side of the cavity were 97\% reflectance, broadband, dielectric mirrors.}
\label{fig:S1}
\end{figure}

\begin{figure}[!h]
\centering\includegraphics[scale=0.9]{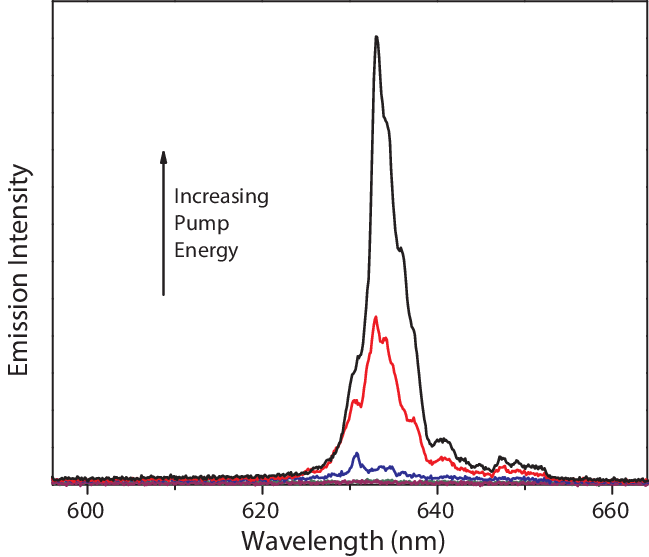}
\caption{Emission spectra from a LiCo laser made by filling FEP tubing with an inner diameter of 1/32" inch with a $0.1\,$wt.\% RhB/glycerol solution. The reflectors on each side of the cavity were 97\% reflectance, broadband, dielectric mirrors.}
\label{fig:S2}
\end{figure}

\clearpage

\begin{figure}[!h]
\centering\includegraphics[scale=1]{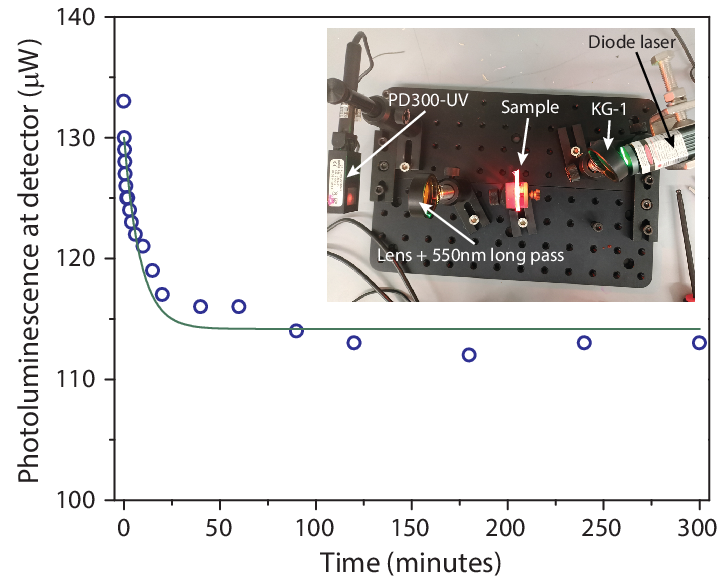}
\caption{Photoluminescence as a function of time for a capillary-filled region of rhodamine B in glycerol at $0.3\,$wt.\% between two glass substrates. The filled region is approximately $2\,$mil thick, $1/4$'' wide, and $3/8$'' long. Experimental data is in circles and the solid line is the fit model. The model assumes a simple rate equation with a constant replenish rate, $dN/dt = -\beta N + N_\mathrm{rep}$. The solution is of the form $N(t) = N_\mathrm{rep}/\beta + \left(N_0 - N_\mathrm{rep}/\beta\right) \mathrm{exp}\left(-\beta t\right)$. The best fit line indicates a rate of approximately $0.12\,$Np/min when illuminated by a $\sim 100\,$mW laser at $532\,$nm with a beam diameter of approximately $1/2\,$cm. The inset shows the setup after the experiment was completed and the room lights were turned back on.}
\label{fig:S3}
\end{figure}

\end{document}